\documentclass[aps,prl,reprint,amsmath]{revtex4-1}

\usepackage{graphicx}
\usepackage{dcolumn}
\usepackage{amsmath}
\usepackage{bm}
\usepackage{float}
\usepackage{color}
\usepackage{amssymb}

\begin{document}

\title{
Current Distribution and Transition Width in Superconducting Transition-Edge Sensors
}

\author{D.S. Swetz}
\email[Electronic mail: ]{swetz@nist.gov}

\author{D.A. Bennett}

\author{K.D. Irwin}

\author{D.R. Schmidt}

\author{J.N. Ullom}

\affiliation{National Institute of Standards and Technology, 325 Broadway MS\,817.03, Boulder, Colorado 80305, USA}

\date{\today}

\begin{abstract}


Present models of the superconducting-to-normal transition in transition-edge sensors (TESs) do not describe the current distribution within a biased TES.  This distribution is complicated by normal-metal features that are integral to TES design.  We present a model with one free parameter that describes the evolution of the current distribution with bias. To probe the current distribution experimentally, we fabricated TES devices with different current return geometries. Devices where the current return geometry mirrors current flow within the device have sharper transitions, thus allowing for a direct test of the current-flow model. Measurements from these devices show that current meanders through a TES low in the resistive transition but flows across the normal-metal features by 40 percent of the normal-state resistance.  Comparison of transition sharpness between device designs reveals that self-induced magnetic fields play an important role in determining the width of the superconducting transition.

\end{abstract}


\maketitle


The sharp change in resistance of a superconductor over a narrow temperature range is both a natural temperature reference and an attractive thermometer.    A Transition-Edge Sensor (TES) consists of a 2-dimensional metal film that is electrically biased into the superconducting phase transition, where its temperature and resistance respond to deposited energy \cite{Irwin:2005}.  TES thermometers have enabled some of the most sensitive calorimetric and bolometric measurements known.  TES measurements of single X-ray, gamma-ray, and alpha quanta achieve the highest resolving powers of any energy-dispersive technique: $E/\triangle E$ $\approx$ 4000-5000 \cite{Smith:2012, Bacrania:2009, Horansky:2008}. Arrays of TES microbolometers are integral to modern submillimeter and millimeter-wave astronomy, achieving microkelvin sensitivity in maps of the cosmic microwave background (for example, see Ref. \onlinecite{Das:2011}).  Despite the broad use and success of these sensors, much remains uncertain about their behavior, including their internal current distribution and the physics that determines the width of the superconducting  transition under bias.  

There are several models for describing the TES transition.  Bennett \textit{et al.} \cite{Bennett:2012} recently extended the two-fluid model of Irwin \textit{et al.} \cite{Irwin:1998} where the TES bias is separated into supercurrent and quasiparticle components. Two-fluid predictions for transition shape were compared to data, but the current fractions were treated as fitting parameters.  Sadleir \textit{et al.} \cite{Sadleir:2010} identified the importance of the proximity effect and treat a TES as a weak link between superconducting leads. Working within the weak-link context, Kozorezov \textit{et al.} \cite{Kozorezov:2011} model a TES as a shunted junction.  However, it is unclear whether a weak-link approach can predict the transition shape under bias, and doubtful whether proximitization from the leads has a strong role in the large devices studied here.  The critical-current variation with field that is the clearest indicator of weak-link behavior is absent in our devices.    All three models lack geometric detail; yet, real-world TESs have additional normal-metal features deposited in complex geometries on the transitioning film.  These features are used to reduce noise and control the transition width \cite{Ullom:2004}, and are likely to influence the flow of the sensor bias current.  

The lack of geometric detail in existing transition models means they cannot treat the effects of self-induced magnetic fields from geometry-dependent bias currents or the effects of field inhomogeneities across a device.  More generally, these approaches assume that the resistive surface is a function only of current and temperature and neglect external or self magnetic fields.  In general, a nonuniform field creates a distribution of critical temperatures and currents within a device that can be expected to broaden the superconducting transition. A crucial unresolved question is the origin of the transition width in a TES, a fundamental parameter that affects application-relevant quantities such as sensor dynamic range and speed.  

In this letter, we map the current distribution in a widely used TES geometry as a function of bias point within the transition.  Further, we show that geometry-specific self-fields affect transition shape and that the transition is significantly broadened by self-fields.  

\begin{figure}
\includegraphics[width=3.4in]{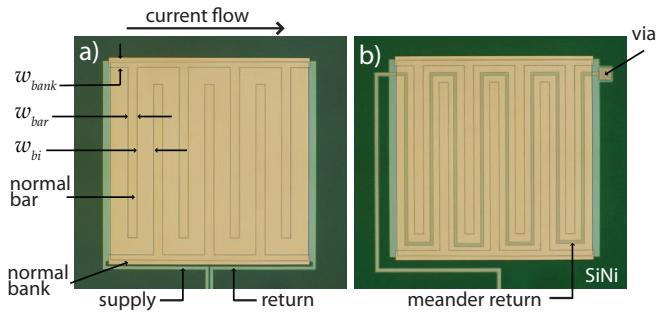}
\caption{\label{fig:sensors}
Pictures of TESs with two bias return geometries.  The base sensors are 350\,$\mu$m by 350\,$\mu$m MoCu bilayers with seven normal-metal bars.  Current flows from left to right.  (a) Standard bias return.  (b) Meander bias return with galvanically isolated superconducting Nb trace deposited over the TES.  For the meander design, the supply and return lines approach the sensor as a microstrip.  For the standard design, the supply and return approach the sensor side-by-side.  The bar width $w_{bar}$ = 16\,$\mu$m, the bank width $w_{bank}$ = 17\,$\mu$m, the bilayer width between bars $w_{bi}$ = 32\,$\mu$m, the bar length $L_{bar}$ = 300\,$\mu$m, the number of normal bars $N_{bar} = 7$, and the bank length $L_{bank}$ = 350\,$\mu$m.}    
\end{figure}

The devices used in this study are representative of TESs used for ultrahigh-resolution calorimetry.  They consist of proximity-coupled layers of Mo and Cu whose thicknesses (100\,nm and 200\,nm, respectively) result in a transition temperature $T_c$ for the whole film of approximately 121\,mK.  The bilayer has planar dimensions of 350\,$\mu$m by 350\,$\mu$m and is deposited on a freestanding silicon-nitride membrane.  An additional 500\,nm Cu layer is deposited on top of the bilayer and patterned by use of standard photolithographic techniques into banks along the bilayer edges parallel to current flow and into seven interdigitated bars perpendicular to current flow.  In our ``standard'' device geometry, shown in Fig. \ref{fig:sensors}a, both the Mo current supply and return lines run along one edge of the TES.

We first calculate the magnitude of the self-fields in our standard design and assess whether it is reasonable that they broaden the superconducting transition.  Consider the self-field in a film of thickness $h$ and width $W$, with the $x$-axis parallel and the $y$-axis perpendicular to the plane of the film.  Solutions for the current density and field profile based on the London equations are given by Rhoderick and Wilson \cite{Rhoderick:1962},  but a uniform current distribution provides a similar result to the expected self-field gradient (see Ref. \onlinecite{Rhoderick:1962}, Fig. 1). For simplicity, we assume a uniform current distribution within the current carrying regions and calculate the perpendicular field component at a point $(x_0, y_0)$ relative to the center of the film using:

\begin{equation}
\label{ByTot}
B_y(x_0,y_0) = \frac{\mu_0 J}{2 \pi (W h)} \int_\frac{-h}{2}^\frac{h}{2} \int_\frac{-W}{2}^\frac{W}{2}  \frac{(x + x_0)}{(x + x_0)^2 + (y_0 - y)^2} dx dy.
\end{equation}

\begin{figure}
\includegraphics[width=3.4in]{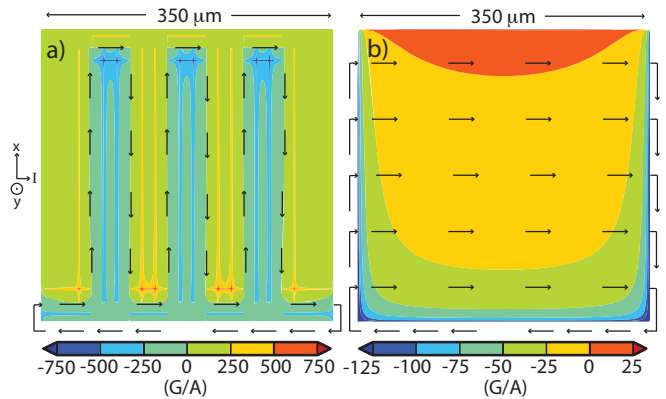}
\caption{\label{fig:totalField}
Estimated field per amp of bias current across our TESs for the standard lead geometry evaluated at the film centerline ($y_0 = 0$; evaluating Eq. \ref{ByTot} away from the $y_0 = 0$ centerline reduces the field per amp by a maximum of $\approx 8\%$ at the film edges $x_0 = \frac{W}{2}$ and $y_0 = \frac{h}{2}$.)  (a) Meandering current pattern.  (b) Uniform current pattern.  The black arrows show the current path.  Typical bias currents range from 60\,$\mu$A to 180\,$\mu$A.}    
\end{figure}

By applying Eq. \ref{ByTot} to our standard device lead geometry, we have calculated the perpendicular field across our sensors, including the field generated by the supply and return leads running along the device edges.  We consider two potential current-flow patterns in the TESs.  One option is for current to meander in a serpentine path between the normal bars of the device.  The second possibility is for current to flow as quasiparticles across the normal bars so that the current density is uniform throughout the TES (horizontally in Fig. \ref{fig:sensors}).  Fig. \ref{fig:totalField} shows the two-dimensional field pattern for both a meandering current and uniform current.  In the meandering case, we model the current as traveling in a 30\,$\mu$m-wide sheet that traverses a serpentine path between the seven bars.  In the uniform case, we assume the current travels in the entire 350\,$\mu$m-wide width of the device.  A meandering current generates field gradients approximately ten times greater than those for a uniform current, and the meandering field pattern is more complex, with several local maxima and minima across the device. Furthermore, the self-field of the meander dominates the field produced by the leads.  However, in the uniform geometry, field from the leads dominates over the TES current contribution.  

In our devices, bias currents range from 180\,$\mu$A to 60\,$\mu$A when the devices are biased between 6\,$\%$ and 60\,$\%$ of the normal-state resistance $R_N$, and are close to 100\,$\mu$A at 20\,$\%$ of $R_N$.  Combining these values with the calculations of Fig. \ref{fig:totalField} result in a predicted self-field variation of 5\,$\mu$T--10\,$\mu$T for the meander case and $\sim$\,0.5\,$\mu$T for uniform current flow. 

The most direct effect of a magnetic field on a sensor is to suppress both its transition temperature and critical current $I_c$.  To quantify this change, we measured the reduction in $T_c$ and $I_c$ as a function of externally driven perpendicular magnetic field \cite{online}.  By combining the measured $T_c$ and $I_c$ suppression with the calculated self-field, we can estimate the contribution of the self-field to the transition width of our sensors.  Transition sharpness is conveniently parametrized by the partial logarithmic derivatives of resistance with respect to temperature at constant current $\alpha_I = (T_0/R_0) \partial R/\partial T \rvert_{I_0}$  and with respect to current at constant temperature $\beta_I = (I_0/R_0) \partial R/ \partial I \rvert_{T_0} $.  For a 5\,$\mu$T field, we measured a $\delta T_c = -0.9$\,mK and a $\delta I_c = -55$\,$\mu$A.  As a first approximation, we imagine that gradients in self-field create a spread in $T_c$ and $I_c$ within a device.  If we assume a linear transition, with $\partial R = R_N$ over the measured $\delta T_c$ and $\delta I_c$, $\alpha_I \sim (T_0/R_0) R_N/\delta T_c$ and $\beta_I \sim (I_0/R_0) R_N/ \delta I_c$.  At a bias of 20\,$\%$, $\alpha_I  \sim$ 670 and $\beta_I  \sim$ 10. These values of $\alpha_I$  and $\beta_I$ are similar to measured results discussed below and presented elsewhere \cite{Jethava:2009}, confirming that self-field magnitudes are sufficient to broaden the superconducting transition \cite{online}.   

To determine the current distribution in a TES, we fabricated devices with sensing elements identical to those of the standard design but with a different return path for the bias current.  This bias-return geometry is shown in Fig. \ref{fig:sensors}b and consists of a ``meander'' return geometry, where a 350\,nm-thick transparent dielectric layer of silicon oxide grown by plasma enhanced chemical vapor deposition, and a via, allow a 300\,nm-thick by 6\,$\mu$m-wide Nb trace to return the bias current on top of the TES in a serpentine pattern between the normal-metal bars.  When current flows in a serpentine path between the normal bars of the device, the meander return geometry produces a field that reduces the perpendicular component of the magnetic field generated by the bias current.  

Since this magnetic field suppresses superconductivity and broadens the superconducting transition, comparison of the transition sharpness between designs can be used to probe the nature of the current distribution.  It is likely that the current flow in the device varies based on the bias point.  We can see this by considering two operating extremes.  When the sensor is low in the resistive transition, the bilayer between the normal-metal bars is nearly perfectly superconducting, and current will meander between the bars.  Higher in the transition, normal resistance will appear in parts of the bilayer.  Once resistance begins to appear, it becomes energetically favorable for current to flow across the bars because of the greater length of the meander path compared to the uniform path, and because the normal resistance of one square of bar material ($R_{\square, bar}$ = $R_{\square, bank}$ = 4.77\,m$\Omega/\square$) is less than the normal resistance of one square of bilayer ($R_{\square, bi}$ = 11.77\,m$\Omega/\square$). 

\begin{figure}
\includegraphics[width=3.3in]{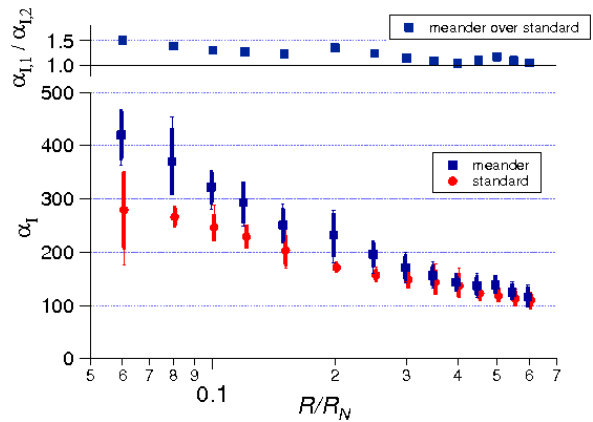}
\caption{\label{fig:alpha_beta}
{Measured $\alpha_I$ versus bias point for the two lead geometries.  Measurements were taken on four different sensors of each bias-return geometry.  Multiple sensors were tested to account for device-to-device variation, and multiple measurements were performed with some sensors.   In total, there are six measurements of $\alpha_I$ at each bias point.  Plot markers indicate average values, thick vertical lines show the standard deviation at each bias point, and thin vertical lines depict the full range of values at each bias point. Markers for the meander geometry taken at the same bias as the standard geometry are slightly offset to the left along the x-axis for clarity.}
}    
\end{figure}

To pursue this hypothesis, we have measured transition sharpness at multiple bias points from 6\,$\%$--60\,$\%$ of $R_N$ corresponding to bias currents from 180\,$\mu$A--60\,$\mu$A, respectively \cite{online}.  Measured values of $\alpha_I$ are shown in Fig. \ref{fig:alpha_beta}.  The measured values of $\beta_I$ show similar trends and are therefore presented in the online supplemental material \cite{online}.  Low in the transition, below about 30\,$\% R_N$, we observe an increase in both $\alpha_I$ and $\beta_I$ for the meander bias return when compared to the standard geometry.  At the lowest bias measured, the average values of both parameters are increased by up to 50\,$\%$ in the meander return as compared to the standard return.  The increase in $\alpha_I$ and $\beta_I$ in the meander geometry demonstrates that current does indeed travel a serpentine path between the bars at low $\% R_N$.      

To understand the results of Fig. \ref{fig:alpha_beta}, we consider the resistance contributions of three parallel current paths through a TES: (1) resistance from the normal banks $R_{banks}$, (2) resistance from a meandering path between the normal bars $R_{meander}$, and (3) resistance from uniform current flow (bus flow) between the normal banks $R_{bus}$.  Following the definitions of the various geometric features given in Fig. \ref{fig:sensors}, the resistance in the banks is given by $R_{banks} = L_{bank}/(2 w_{bank}) R_{\square, bank}$.  $R_{meander}$ depends on the strength $f$ of superconductivity in the bilayer, where $f$ ranges from 0 to 1, corresponding to fully superconducting  and  fully normal.   For the devices of Fig. \ref{fig:sensors}, $R_{meander} = \{2+w_{bar}/w_{bi}+(N_{bar}-1)(1+ w_{bar}/w_{bi} + L_{bar}/w_{bi})\}f R_{\square, bi}$.  $R_{bus}$ depends on the distance in the bilayer $\Lambda_{Q^{\ast}}$ over which a normal current that has traversed a bar returns to the condensate as a supercurrent.  The quantity $\Lambda_{Q^{\ast}}$ is the well known branch-imbalance length \cite{Tinkham:2003}. Here, $R_{bus} \approx N_{bar}\{w_{bar} R_{\square, bar}/L_{bar} + 2 \Lambda_{Q^{\ast}}R_{\square, bi}/(L_{bar}+w_{bi})\}  + f R_{\square, bi}/(L_{bar}+w_{bi}) \{2\,(w_{bi}-\Lambda_{Q^{\ast}}) + (N_{bar}-1)(w_{bi}-2 \Lambda_{Q^{\ast}})\}$.  We take device resistance $R$ to be the parallel combination of $R_{banks}$, $R_{meander}$, and $R_{bus}$; when $f$=1, the device is normal with a resistance $R =R_N$.  A schematic illustrating the meander and uniform current paths in the device and algebraic derivations of $R_{meander}$ and $R_{bus}$ are given in the online supplemental material \cite{online}.  

\begin{figure}
\includegraphics[width=3.4in]{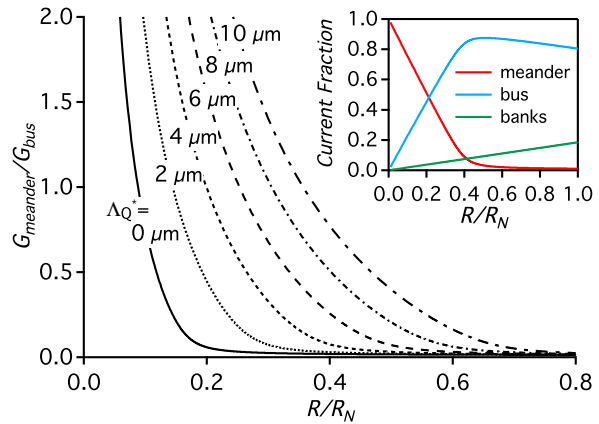}

\caption{\label{fig:CurrentDistribution}
{Calculated ratio of $G_{meander}/G_{bus}$ versus bias point as a function of $\Lambda_{Q^{\ast}}$.  The measurements of transition sharpness (Fig. \ref{fig:alpha_beta} and Ref. \cite{online}, Fig. 1) indicate that $G_{meander} \gtrsim G_{bus}$ for bias values below 0.3 $R/R_N$, but that $G_{meander} \ll G_{bus}$ above biases of 0.4 $R/R_N$.  These conditions constrain $\Lambda_{Q^{\ast}}$ to a value between 3\,$\mu$m and 6\,$\mu$m, where $G_{meander}/G_{bus} \gtrsim 0.25$ for $R/R_N$ below 0.3 and $G_{meander}/G_{bus} \lesssim 0.25$ for $R/R_N$ above 0.4.  \textit{Inset.} Calculated fraction of current in the three current paths versus bias for $\Lambda_{Q^{\ast}}$ = 4.5\,$\mu$m.}
}    
\end{figure}

From the expressions above, we expect a cross-over resistance $R_X$ that depends on $\Lambda_{Q^{\ast}}$ above (below) which $G_{meander} = R_{meander}^{-1}$ is less (greater) than $G_{bus} = R_{bus}^{-1}$. Values of $G_{meander}/G_{bus}$ are plotted versus $R/R_N$ for varying $\Lambda_{Q^{\ast}}$ in Fig. \ref{fig:CurrentDistribution}.  Our measurements of transition sharpness for the meander and standard geometries show that significant current flows in a meander pattern for $R/R_N$ as high as 0.3.  Therefore, $G_{meander} \gtrsim G_{bus}$ low in the transition below 0.3\,$R/R_N$.  At $R/R_N \geq 0.4$, there is no measurable difference in the transition sharpness between the two device geometries, indicating that the majority of the current is in the bus mode, hence $G_{meander} \lesssim G_{bus}$ at biases above 0.4\,$R/R_N$.  Low in the transition, we take the constraint that $G_{meander}/G_{bus} \gtrsim 0.25$.  Similarly, high in the transition we take the constraint that $G_{meander}/G_{bus} \lesssim 0.25$.  These conditions are met by the curves in Fig. \ref{fig:CurrentDistribution}, where $\Lambda_{Q^{\ast}}$ falls between 3 and 6\,$\mu$m.  We note that while our choice of constraints on $G_{meander}/G_{bus}$ is subjective, varying these constraints only shifts the bounds on $\Lambda_{Q^{\ast}}$.  For $\Lambda_{Q^{\ast}}$ = 4.5\,$\mu$m (used hereafter), $R_X/R_N =$ 0.21.  We make several observations.  First, measurements of a macroscopic parameter, transition sharpness, provide remarkably tight constraints on the microscopic parameter $\Lambda_{Q^{\ast}}$.  Second, these results provide clear proof of the importance of quasiparticle transport in TES bilayers under realistic working conditions, since $\Lambda_{Q^{\ast}} =$ 0 is inconsistent with the data.  Third, because the meander path length is greater than the uniform path length and because $R_{\square, bar} < R_{\square, bi}$, $R_X$ is achieved at a very low value, $f_X$ = 0.005, indicating that the appearance of almost infinitesimal resistance values in the bilayer will begin the transfer of current into the bus mode.  




Having determined $\Lambda_{Q^{\ast}}$, we can compute the fraction of the device current in the meander and bus modes, as well as in the banks.  A current fraction $i$ is given by $G_i/(G_{bus}+G_{banks}+G_{meander})$.  The three current fractions are plotted versus $R/R_N$ in the inset of Fig. \ref{fig:CurrentDistribution}.  At low $R/R_N$, most current is in the meander, but current in the bus mode increases with $R/R_N$, and the two modes are equal at $R_X/R_N$ = 0.21.

The large difference in transition sharpness between device geometries sheds light on an important unresolved question, namely the origin of the transition width in biased TES devices.  Not only can the relative values of $\alpha_I$ and $\beta_I$ be used to locate current flow, the quantitative differences in $\alpha_I$ and $\beta_I$ indicate that self-fields play an important role in determining the transition width.  Since transition width is proportional to $1/\alpha_I$, the transition in the meander geometry is roughly 2/3 narrower than in the standard geometry.  Because the designs differ only in the self-field experienced by the TES, as much as 1/3 of the transition width in the standard device geometry at realistic bias points between 0.05--0.2\,$R_N$ must be due to self-field effects.  

The calculated self-fields of Fig. \ref{fig:totalField} provide additional insight into the measurements of Fig. \ref{fig:alpha_beta}.  Above $R_X$, the current flow switches from a meandering supercurrent to uniform current flow.  At high bias, we might expect that the standard design would have a sharper transition than that of the meander design, as the fields from the device current and the meander return no longer cancel.  However, the current is decreasing, and the total field gradient across the sensor also drops dramatically, as shown in Fig. \ref{fig:totalField}b, so self-field effects simply become less relevant.  

In summary, our findings provide strong evidence that device geometry, particularly normal-metal features, influences the flow of current in a TES.  For a sensor with interdigitated normal-metal bars, we have shown that current travels in a serpentine path between the bars at low $\% R_N$, but that this is completely finished by 40\,$\%$ of $R_N$.  We have developed a model of the current distribution as a function of device bias and used this model to constrain the quasiparticle branch imbalance length $\Lambda_{Q^{\ast}}$.  This bias current produces an inhomogeneous self-induced magnetic field across the sensor.  These self-fields contribute to the superconducting transition width.  Low in the transition, the $\alpha_I$ and $\beta_I$ differences between the meander and standard current returns indicate that self-field effects can account for as much as 1/3 of the observed transition widths.  Because TES calorimeters are typically biased between 5$\%$ and 20$\%$ of $R_N$, self-field effects are most relevant at actual working conditions.  Our measurements of the transition broadening from self-fields are likely a lower bound, since several effects contribute to imperfect field cancellation by the meander return: the 350\,nm thickness of the dielectric, the somewhat arbitrary 6\,$\mu$m width of the Nb meander, and a layout error where the meander return enters and exits the TES at the top of the device (as shown in Fig.\ref{fig:sensors}) but should enter and exit at the bottom.

These results provide insight into the physics governing the superconducting transition in 2-dimensional films with additional normal-metal features.  Our determination of the internal current distribution may inform future models that move beyond a strictly monolithic picture of TES devices.  While we have shown that self-fields account for a significant fraction of the transition width at low $\% R_N$, other effects must also play a role in determining the transition width.  A likely mechanism is spatial $T_c$ and $I_c$ variation within the device induced by the lateral proximity effect between the bilayer and normal-bar regions.  Finally, we speculate that the observed suppression of excess noise by normal-metal features \cite{Ullom:2004} is because they rapidly force current into a quasiparticle mode that resembles current flow in conventional resistors and is less susceptible to fluctuation mechanisms associated with the superconducting phase transition.  

We gratefully acknowledge support from the NASA APRA program.  D. Swetz was supported in part by a NRC fellowship.  Contribution of NIST, not subject to copyright.

%

\end{document}